\documentclass{nature}

\usepackage{hyperref}
\usepackage{epsfig}
\usepackage{xcolor}
\usepackage{lscape}
\usepackage[acronym]{glossaries}
\usepackage{tablefootnote}
\usepackage{amssymb}
\usepackage{pdfpages}

\newcommand*\degr{\ensuremath{^\circ}}
\newcommand*\arcmin{\ensuremath{^\prime}}

\newcommand*\sun{\ensuremath{\odot}}

\title{A rich population of free-floating planets in the Upper Scorpius young stellar association}

\author{Núria Miret-Roig$^{1,2}$ \& Hervé Bouy$^{1}$ \& Sean~N.~Raymond$^{1}$ \& Motohide Tamura$^{3,4}$ \& Emmanuel Bertin$^{5,6}$ \& David Barrado$^{7}$ \& Javier Olivares$^{1}$ \& Phillip~A.~B. Galli$^{1}$ \& Jean-Charles Cuillandre$^{8}$ \& Luis Manuel Sarro$^{9}$ \& Angel Berihuete$^{10}$ \& Nuria Huélamo$^{7}$}

\begin{document}

\maketitle

\begin{affiliations}
 \item Laboratoire d'Astrophysique de Bordeaux, Univ. Bordeaux, CNRS, B18N, all{\'e}e Geoffroy Saint-Hilaire, 33615 Pessac, France
 \item University of Vienna, Department of Astrophysics, Türkenschanzstraße 17, 1180 Wien, Austria; e-mail: \url{nuria.miret.roig@univie.ac.at}
\item Department of Astronomy, Graduate School of Science, The University of Tokyo, Tokyo, Japan
\item Astrobiology Center, National Institutes of Natural Sciences, Tokyo, Japan
\item CNRS, UMR 7095, Institut d’Astrophysique de Paris, F-75014 Paris, France
\item Sorbonne Université, Institut d’Astrophysique de Paris, F-75014 Paris, France
\item Centro de Astrobiología (CSIC-INTA), Depto. de Astrofísica, ESAC Campus, Camino Bajo del Castillo s/n, 28692, Villanueva de la Cañada, Madrid, Spain
\item AIM, CEA, CNRS, Université Paris-Saclay, Université de Paris, F-91191 Gif-sur-Yvette, France
\item Depto. de Inteligencia Artificial, UNED, Juan del Rosal 16, 28040 Madrid, Spain
\item Depto. Estadística e Investigación Operativa, Universidad de Cádiz, Avda. República Saharaui s/n, 11510 Puerto Real, Cádiz, Spain
\end{affiliations}

%% ACRONYMS
\newacronym{BIC}{BIC}{Bayesian information criterion}
\newacronym[plural={FFPs}, \glsshortpluralkey={FFPs}]{FFP}{FFP}{free-floating planet}
\newacronym[plural={GMMs}, \glsshortpluralkey={GMMs}]{GMM}{GMM}{Gaussian mixture model}
\newacronym{KDE}{KDE}{kernel density estimation}
\newacronym{oph}{Oph}{Ophiuchus}
\newacronym{USC}{USC}{Upper Scorpius} 
\newacronym{CADC}{CADC}{Canadian Astronomy Data Centre}
\newacronym{CDS}{CDS}{Centre de Données astronomiques de Strasbourg} 
\newacronym{CFHT}{CFHT}{Canada-France-Hawaii Telescope}
\newacronym{CTIO}{CTIO}{Cerro Tololo Inter-American Observatory}
\newacronym{DANCe}{DANCe}{Dynamical Analysis of Nearby ClustErs}
\newacronym{DECam}{DECam}{Dark Energy Camera}
\newacronym{ESO}{ESO}{European Southern Observatory} 
\newacronym{GDR2}{{\it Gaia} DR$2$}{\emph{Gaia} Data Release $2$}
\newacronym{HSC}{HSC}{Hyper Suprime-Cam} 
\newacronym{ING}{ING}{Isaac Newton Group}
\newacronym{INT}{INT}{Isaac Newton Telescope}
\newacronym{KPNO}{KPNO}{Kitt Peak National Observatory} 
\newacronym{NAOJ}{NAOJ}{National Astronomical Observatory of Japan}
\newacronym{NOAO}{NOAO}{National Optical Astronomy Observatory}
\newacronym{NOIRLab}{NOIRLab}{National Optical-Infrared Astronomy Research Laboratory}
\newacronym{PTF}{PTF}{Palomar Transient Factory}
\newacronym{SMOKA}{SMOKA}{Subaru‐Mitaka‐Okayama‐Kiso‐Archive}
\newacronym{WFC}{WFC}{Wide Field Camera} 
\newacronym{WSA}{WSA}{WFCAM Science Archive}

\begin{abstract}
\Glspl{FFP} are planetary-mass objects that are not bound to host stars. First discovered in the 1990s, their nature and origin are still largely unconstrained because of a lack of large homogeneous samples enabling a statistical analysis of their properties. To date, most \glspl{FFP} have been discovered using indirect methods; micro-lensing surveys have proven particularly successful to detect these objects down to a few Earth masses\cite{Mroz+2020,2021AJ....161..126R}. However, the ephemeral nature of micro-lensing events prevents any follow-up observations and individual characterisation. Several studies have identified \glspl{FFP} in young stellar clusters\cite{Scholz+2012b,Pena-Ramirez+2012} and the Galactic field\cite{Mroz+2017} but their samples are small or heterogeneous in age and origin. Here we report the discovery of between 70 and 170 \glspl{FFP} (depending on the assumed age) in the region encompassing \gls{USC} and \gls{oph}, the closest young OB association to the Sun. It is the largest homogeneous sample of nearly coeval \glspl{FFP} discovered to date. We found an excess of \glspl{FFP} by a factor of up to seven compared to core-collapse models predictions\cite{Chabrier05, Haugbolle+2018, Bate+2019}, demonstrating that other formation mechanisms may be at work. We estimate that ejection from planetary systems might have a contribution comparable to that of core-collapse in the formation of \glspl{FFP}. Therefore, ejections due to dynamical instabilities in giant exoplanet systems must be frequent within the first 10~Myr of a system's life.

\end{abstract}

To date, most exoplanets have been detected through radial velocity or photometric modulations induced in their host stars\cite{Charbonneau+2000, Lissauer+2011, Howard+2012, Mayor+2011, Howard+2010}. As such, the vast majority of known exoplanets are gravitationally bound to stars. However, several \acrfullpl{FFP} have been discovered over the last two decades in astro-photometric surveys of nearby star-forming regions\cite{Scholz+2012b,Pena-Ramirez+2012,Luhman+2016,Esplin+2017, Zapatero-Osorio+2017,Lodieu+2018,Esplin+2019}, young associations\cite{Liu+2013,Kellogg+2015,Schneider+2016,Best+2017}, the solar neighbourhood\cite{Kirkpatrick+2019,Kirkpatrick+2021} and in gravitational micro-lensing surveys of the Galactic field\cite{Mroz+2017}. These ultra-faint objects are incapable of sustaining nuclear fusion and steadily fade in time, making them easier to observe when they are very young. \glspl{FFP} are compact objects of less than about $13$ Jupiter masses that are not bound to a star or brown dwarf but rather wander among them. At present, four scenarios have been proposed for the formation of these extreme objects: a) a scaled-down version of star formation via core-collapse\cite{Padoan+2002, Hennebelle+2008}; b) within a protoplanetary disc, either like gas-giant planets through core accretion\cite{1996Icar..124...62P} or like companions through gravitational fragmentation of massive extended discs\cite{1998EM&P...81...19B, Bate+2002}, followed by ejection by dynamical scattering between planets in both cases\cite{Veras+2012}; c) as aborted stellar embryos ejected from a stellar nursery before the hydrostatic cores could build up enough mass to become a star\cite{Reipurth+2001} and d) through the photo-erosion of a prestellar core by stellar winds from a nearby OB star\cite{Whitworth+2004}. While direct observational evidence confirms that these different processes are all at work\cite{Testi+2016, Fontanive+2020}, we still do not understand their relative contributions to the overall \gls{FFP} population.

Here we present a search for \glspl{FFP} in the 171~deg$^2$ region occupied mainly by \acrfull{USC} and \acrfull{oph}. We selected an elliptical area centred in (RA $=243.5$\degr, Dec $=-23.1$\degr) with a semi-major axis of $8.5\degr$ in RA and a semi-minor axis of $6.4\degr$ in Dec where the spatial and temporal coverage of the observations is the best. This large complex represents the perfect hunting ground to search for young and nearly coeval \glspl{FFP} thanks to its proximity (120--145~pc) and youth (1--10~Myr\cite{Greene+1995, Sullivan+2021, David+2019, Pecaut+2016}).
We combined our ground-based observations in the optical and infrared with wide-field images available in various public archives (see methods). We processed and analysed a total of 80\,818 individual wide-field images acquired with  18 different cameras over the past 20 years to obtain a final catalogue, the \gls{DANCe}\cite{Bouy+13} catalogue, containing positions, proper motions and multi-wavelength photometry ($grizyJHK$) for more than 26 million objects. We complemented the DANCe catalogue with the astrometry and photometry of the \gls{GDR2}\cite{GaiaColBrown+18} and \textit{Hipparcos}\cite{Brown+1997} catalogues. We used this dataset to compute membership probabilities to \gls{USC} and \gls{oph} using a probabilistic model of the distribution of the observable quantities in both the cluster and background populations (see methods). To identify the \glspl{FFP} we used the parameter space that contains the largest amount of parameters and sources with complete information, namely proper motions and $iJHK$ photometry.

\begin{figure}
  \begin{center} 
    \includegraphics[width=0.9\textwidth]{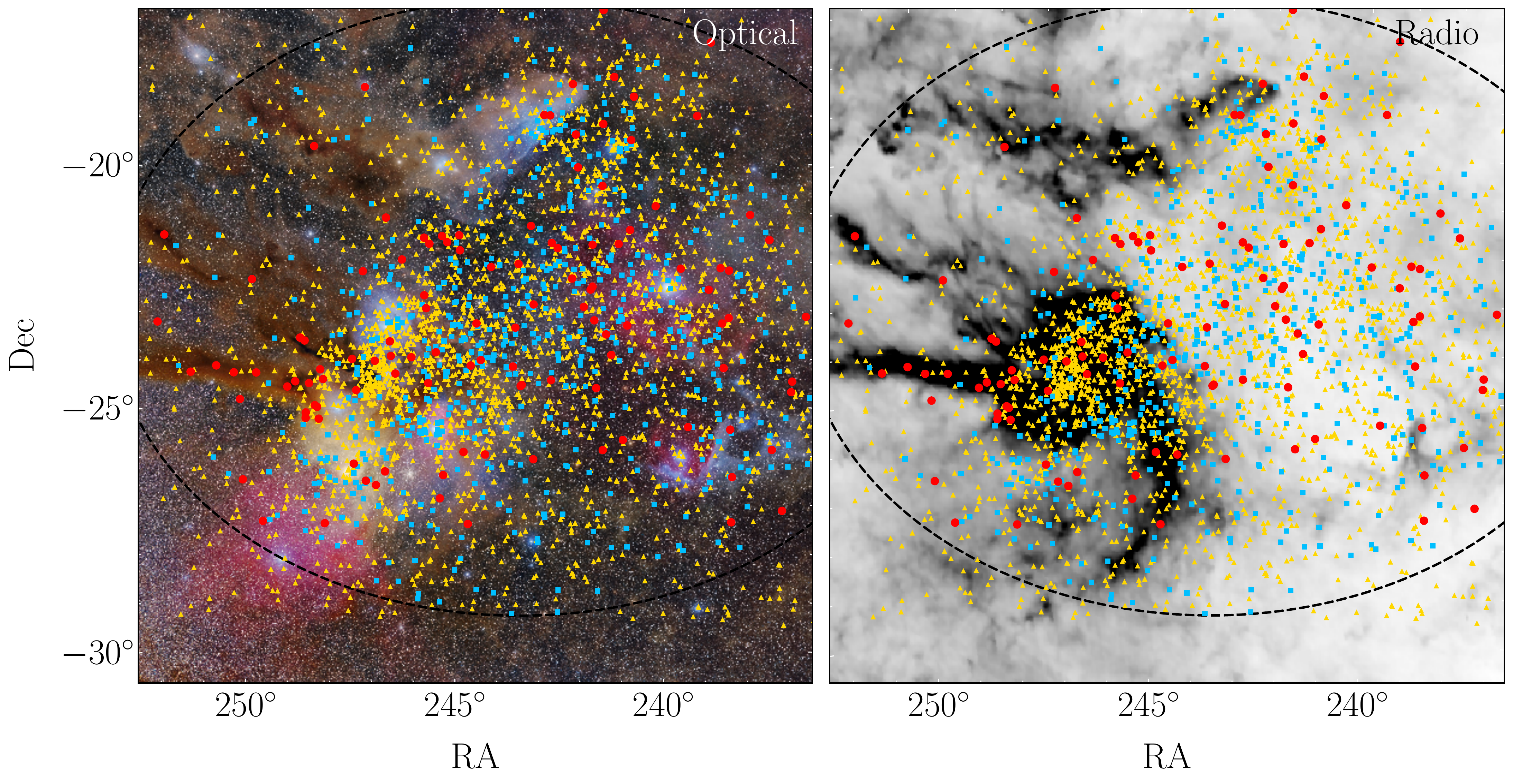}
    \caption[]{Sky distribution of stars (gold triangles), brown dwarfs (blue squares), and \glspl{FFP} (red dots) discovered in this study and classified assuming an age of 5 Myr. The dashed ellipse indicates the area analysed with the DANCe catalogue (see methods). The background images are in the optical (credit: Mario Cogo\cite{Cogo2017}) and at 857~GHz (credit: \textit{Planck}\cite{Planck+2020}).} 
     \label{fig:sky_distribution}
    \end{center}
\end{figure}

We identified 3\,455 high probability candidate members in the area covered by our study, including between 70 and 170 \glspl{FFP} depending on the age assumed for the region. Approximately 20\% of the members are new compared to previous studies\cite{Muzic+2012,Ducourant+2017,Esplin+2018,Damiani+2019,Lodieu+2018,Lodieu+2021}, and this proportion increases to 75\% in the planetary mass domain. It is the largest and most homogeneous sample of \glspl{FFP} identified using direct images so far, and it constitutes an excellent benchmark to test star and planet formation theories. Figure~\ref{fig:sky_distribution} shows the area covered by our survey in the optical and radio wavelengths. The members are distributed all over the region. The majority of members (and \glspl{FFP}) lay in the area of \gls{USC}, where the extinction by the interstellar medium is lower ($A_V<1$~mag). 

\begin{figure}
  \begin{center}
  \includegraphics[width=0.9\textwidth]{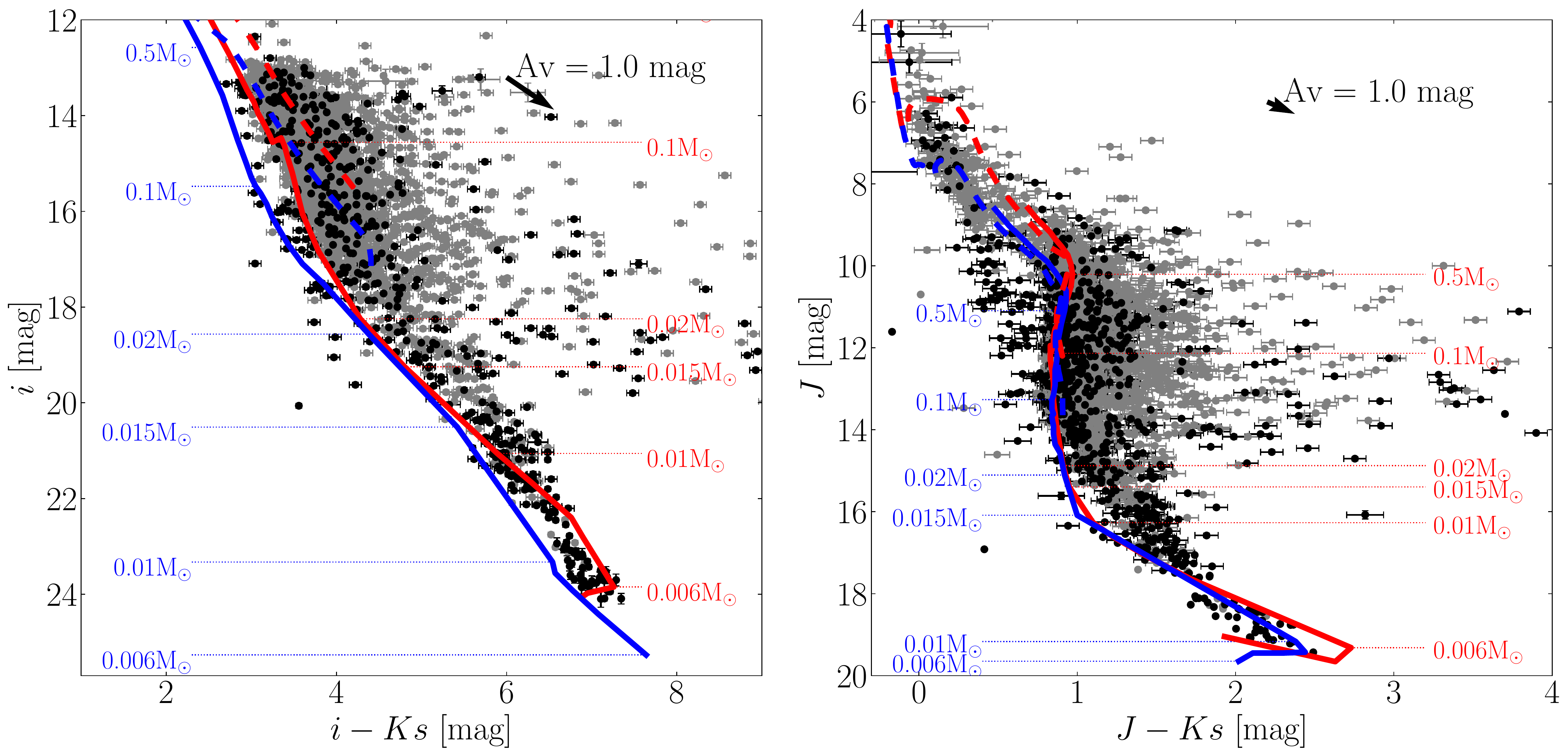}
    \caption[]{Colour-magnitude diagram of the members of \gls{USC} and \gls{oph} identified in this work: previously known members (gray) and new members (black). The error bars represent the uncertainty in the photometry reported in the \textit{Gaia} and DANCe catalogues. The BHAC15 isochrones\cite{Baraffe+15} (solid lines) and the PARSEC-COLIBRI isochrones\cite{Marigo+17} (dashed lines) at 3~Myr (red) and 10~Myr (blue) as well as the extinction vector are overplotted. } 
     \label{fig:CMD}
    \end{center}
\end{figure}

We used our comprehensive membership analysis to study the origin of the \gls{FFP} population in \gls{USC} and \gls{oph}. We computed the number density of members as a function of their masses. This “mass function” of the region constitutes a fundamental constraint for formation theories because different mechanisms predict different relative abundances of stellar, sub-stellar and planetary-mass objects. The transformation from observed luminosities to masses requires knowing the age and using evolutionary models. The age spread of the region\cite{Greene+1995, Sullivan+2021, David+2019, Pecaut+2016}, as well as the complex and overlapping spatial and kinematic distributions of the different coexisting populations, prevented us from disentangling the various groups and assigning ages to individual objects. Instead, we chose to assign ages of 3, 5 and 10~Myr for the entire sample and assumed that the underlying real mass distribution must be included between these borderline cases. We used the entire spectral energy distribution available for each object and the theoretical evolutionary models\cite{Baraffe+15,Marigo+17} to infer the individual mass of each member (see methods). Figure~\ref{fig:CMD} shows colour-magnitude diagrams of the members in two different spaces. We overplotted the theoretical isochrones at the extreme ages mentioned above to illustrate the uncertainties in the mass inherited from the lack of individual precise ages.

\begin{figure}
  \begin{center} 
  \includegraphics[width=0.85\textwidth]{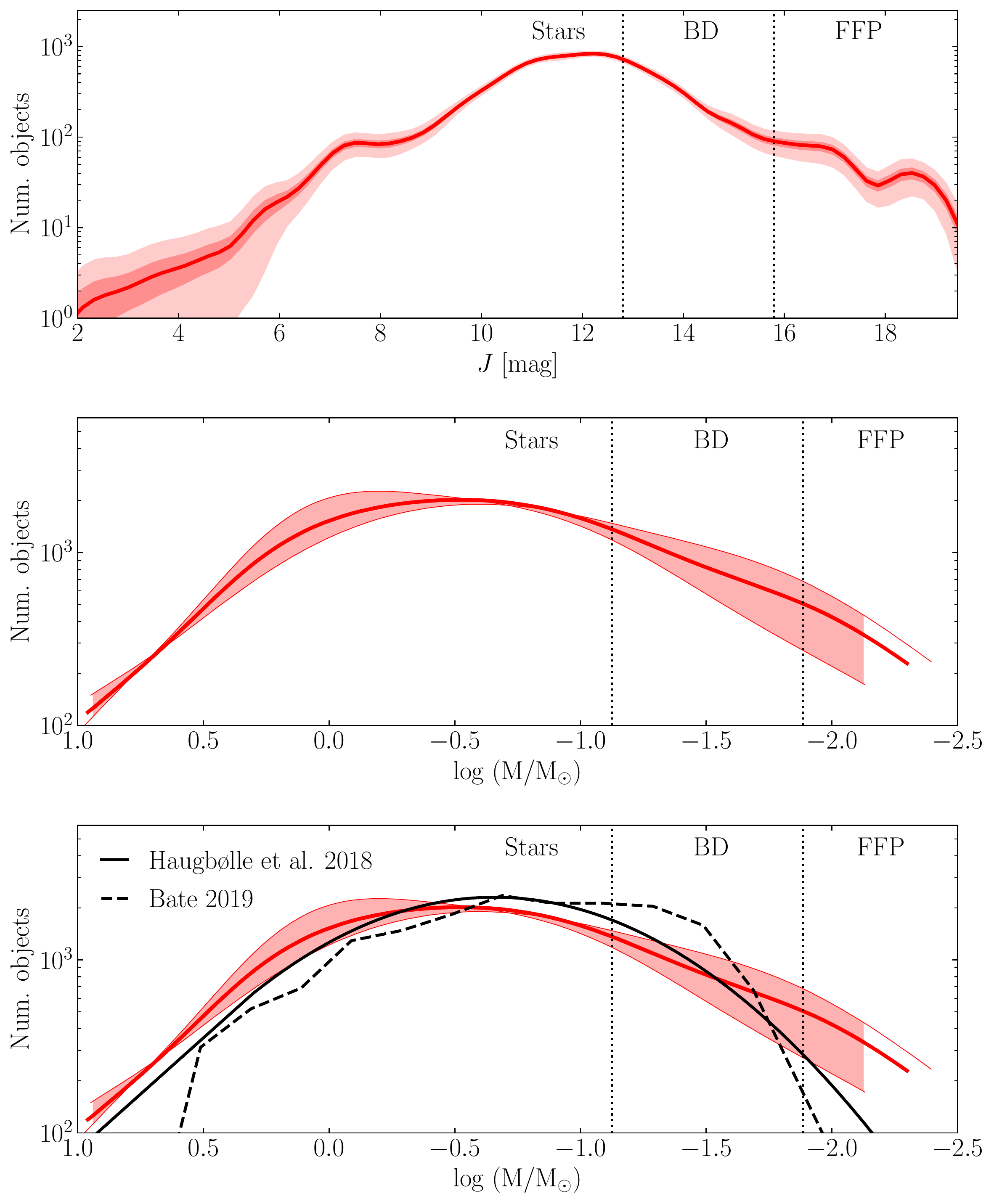}
    \caption[]{$J$ apparent magnitude distribution (top) and mass function (middle and bottom) of the members of \gls{USC} and \gls{oph}. The shaded regions indicate the $1$ and $3\sigma$ uncertainties from a bootstrap (top) and the dispersion due to the age (3--10~Myr, middle and bottom). The mass functions from simulations\cite{Haugbolle+2018, Bate+2019} are overplotted on our observational mass function (bottom). All the functions are normalised in the mass range 0.004--10~M$_\sun$. The hydrogen  ($75$~M$_{Jup}$) and deuterium ($13$~M$_{Jup}$) burning limits are indicated by the vertical dotted lines according to the BHAC15 evolutionary models\cite{Baraffe+15} and assuming an age of 5~Myr.  } 
     \label{fig:mag-mass-func}
    \end{center}
\end{figure}

The $J$ apparent magnitude distribution (Fig.~\ref{fig:mag-mass-func}, upper panel) is a direct product of the observations, and thus, it is not affected by the uncertainties and errors in the transformation from luminosities to masses. Thanks to improved statistics achieved with our larger sample we unambiguously identified a dip at planetary masses (apparent magnitude $J\sim17.8$~mag, corresponding to absolute magnitude $M_J\sim12$~mag and masses 7--13~M$_{Jup}$ depending on the age assumed) which was also reported in volume-limited samples in the solar neighbourhood\cite{Bardalez-Gagliuffi+2019,GaiaColSmart+2021}. Are we seeing changes relative to the formation mechanisms? Could this be the “real” frontier between brown dwarfs and \glspl{FFP}? To our knowledge, the origin of this dip remains unknown and more studies are needed to answer these questions. However, the presence of this feature at the young ages of \gls{USC} and \gls{oph} (1--10 Myr) suggests that it must be the result of the formation and/or early evolution of these objects. 

Figure~\ref{fig:mag-mass-func} (middle panel) shows the mass function at 5~Myr (which can be considered an intermediate age of the entire complex); the mass functions at 3 and 10~Myr define the upper and lower confidence intervals. A number of details and features clearly significant in the magnitude distribution (in particular the dip mentioned above) are blurred and lost in the mass distribution likely due to the uncertainties related to the transformation from luminosity to mass. 
Our magnitude and mass distributions show a rich population of \glspl{FFP}. We evaluated the contamination rate in this sample, mostly due to background reddened giant stars and background Galaxies, using two different methods. First, using synthetic data\cite{Olivares+19}, we estimated a contamination rate in the planetary mass domain smaller than $4\%$. Second, using sources for which both DANCe and \gls{GDR2} measurements are available (in the range $9<J<14$~mag, hence above the planetary mass regime) and assuming that the \gls{GDR2} sample represents the ground truth, we estimated a contamination rate of approximately 8\%. The real contamination rate of \glspl{FFP} is likely in between these two values and should be confirmed by follow-up spectroscopic observations. The number of \glspl{FFP} reported in our analysis must constitute a lower limit of the actual total number of \glspl{FFP} since our analysis is expected to miss the objects most affected by extinction ($A_{\rm V}\gtrsim3$~mag), as well as objects displaying a large near-infrared excess related to the presence of circumstellar material. 

The fraction of \glspl{FFP} in our sample, meaning the relative proportion of \glspl{FFP} to stars and brown dwarfs, is $0.045^{+0.023}_{-0.029}$, where the uncertainties come from the uncertainty on individual ages (between 3 and 10 Myr). We estimated this fraction by first integrating the observational mass function (Fig.~\ref{fig:mag-mass-func}, middle panel) in the \gls{FFP} (4--13~M$_{Jup}$), brown dwarf (13--75~M$_{Jup}$) and stellar (0.075--10~M$_\sun$) mass regimes and then computing the ratio between \glspl{FFP} and brown dwarfs plus stars. While previous studies reported a similar fraction of \glspl{FFP}\cite{Scholz+2012b, Pena-Ramirez+2012}, our sample doubles the absolute number of \glspl{FFP} in a single association, which significantly reduces the statistical uncertainties in the mass function. A recent photometric study in the central 6~deg$^2$ region of \gls{USC}\cite{Lodieu+2021} found a ratio of 1.0--1.5 planetary-mass members per square degree, this result is broadly consistent within the statistical uncertainties with the ratio we find in the 171~deg$^2$ area covered by our survey (0.5--1.0 planetary-mass members per square degree).  

In the high mass regime ($>1$~M$_\sun$), our mass function has a slope of $\Gamma =-1.2\pm0.2$  (measured in units of logarithmic mass), compatible with the Salpeter slope\cite{Salpeter1955} and with all the models and simulations considered here\cite{Chabrier05, Thies+07, Thies+15, Haugbolle+2018, Bate+2019}. In the substellar mass regime ($<75$~M$_{Jup}$), our observational mass function (Fig.~\ref{fig:mag-mass-func}, bottom) has a slope of $\Gamma =0.62^{+0.13}_{-0.01}$ (measured in units of logarithmic mass), similar to values reported in the field population for L, T and Y dwarfs\cite{Kirkpatrick+2019, Kirkpatrick+2021} and to models including several channels of substellar object formation\cite{Thies+07,Thies+15}. The mass function over the low-mass stars and high-mass brown dwarfs regime (0.03--1~M$_\sun$) is compatible with a log-normal distribution\cite{Haugbolle+2018, Chabrier05}. When integrating the analytical mass function of models including mostly core-collapse formation \cite{Chabrier05,Haugbolle+2018} over the planetary mass range (4--13~M$_{Jup}$), we find that they predict a fraction of only $0.009-0.019$ \glspl{FFP}, underestimating up to seven times our measurement (depending on the age assumed). 
This excess of \glspl{FFP} with respect to a log-normal mass distribution is in good agreement with the results reported in $\sigma$~Orionis\cite{Pena-Ramirez+2012}. 
Interestingly, our observational mass function also has an excess of low-mass brown dwarfs and \glspl{FFP} with respect to simulations including both core-collapse and disc fragmentation\cite{Bate+2019}. This suggests that some of the \glspl{FFP} in our sample could have formed via fast core-accretion in discs rather than disc fragmentation. We also note that the continuity of the shape of the mass function at the brown dwarf/planetary mass transition suggests a continuity in the formation mechanisms at work for these two classes of objects.

Hereafter, we use the current knowledge of planet and star formation to discuss the origin of \glspl{FFP}. The fraction of observed \glspl{FFP} ($f_\textup{FFP~observed}$) is the sum of \glspl{FFP} formed by ejection from a disc ($f_\textup{FFP~ejected}$), the \glspl{FFP} formed by core-collapse ($f_\textup{FFP~core~collapse}$) and the \glspl{FFP} formed by other mechanisms ($f_\textup{FFP~other}$, including photo-evaporation and ejection from a prestellar cluster). The fraction of \glspl{FFP} ejected from a planetary system $(f_\textup{FFP~ejected})$ depends on the fraction of stars and brown dwarfs that form giant planets ($f_\textup{giant}$), on the fraction of such planetary systems that become unstable ($f_\textup{unstable}$), and on the number of ejected planets per unstable system ($n_\textup{ejected}$). This formulation is similar to that of previous studies\cite{Veras+2012} and has the following expression in our study, for objects more massive than 4~M$_{Jup}$.
\begin{equation}
\begin{aligned}
    f_\textup{FFP~observed} &= f_\textup{FFP~ejected} + f_\textup{FFP~core~collapse} + f_\textup{FFP~other} \\
    &= f_\textup{giant}\cdot f_\textup{unstable}\cdot n_\textup{ejected} + \int_{4~\textup{M}_{Jup}}^{13~\textup{M}_{Jup}} \xi_\textup{core~collapse}(m)~dm + f_\textup{FFP~other}
    \label{eq:FFP}
\end{aligned}
\end{equation}

The fraction of stars that forms giant planets ($f_\textup{giant}$) is constrained by the observed demographics of giant exoplanets measured by radial velocity, direct-imaging, transit and micro-lensing surveys\cite{Fernandes+2019, Clanton+2017C, Bowler2016, Suzuki+2016, Mayor+2011, Howard+2012, Fressin+2013, Wittenmyer+2020, Cumming+2008, Butler+2006}. The planetary occurrence rate depends on many astrophysical parameters (the host star and planet masses, orbital separation, stellar metallicity, and others), and each of the techniques above mentioned is sensitive to a specific region of the parameter space\cite{Winn+2015}. Therefore, we combined the occurrence rates obtained with different techniques to minimise the possible observational biases. In Table~\ref{tab:POR}, we summarise the occurrence rates obtained by different authors and describe the properties of each study.
The fraction of planetary systems that become unstable ($f_\textup{unstable}$) must be at least 75\% to match the observed distribution of giant planet eccentricities\cite{Juric+2008, Chatterjee+2008, Raymond+2010, Ford+2008, Ida+2013}. A minimum of two planets per system is needed for instability to happen, and indeed, many giant exoplanets are found in multiple planetary systems or contain hints (such as radial velocity trends) of additional companions\cite{Cumming+2008}. The number of ejected planets per unstable system ($n_\textup{ejected}$) scales with the number of planets involved in the instability\cite{Veras+2012}. We consider a simplified scenario in which planetary systems contain two to four giant planets, the three cases being equally likely, and every time a system becomes unstable it ejects one planet. With these assumptions, we find $n_\textup{ejected}\sim1/3\cdot2\cdot1/2 + 1/3\cdot3\cdot1/3 +1/3\cdot4\cdot1/4 \sim1$.

Combining the upper and lower limits on the three factors defining the fraction of \glspl{FFP} ejected from planetary systems, we obtain $f_\textup{FFP~ejected}\sim 0.005-0.021$, which represents between 10--130\% of the \gls{FFP} population we found. The large uncertainty in this percentage is due to the lack of precise masses in our observations on the one hand (which itself is related to the lack of precise individual ages) and to the uncertainty on the occurrence rate of planets and their ejection process on the other hand. A percentage of planets formed by ejection above 100\% is obviously of no physical meaning and only reflects the limitations of our assumptions and simplifications, which are based on the best current knowledge of planetary systems and the latest evolutionary models. This result nevertheless suggests that ejection from planetary systems is a significant mechanism for \glspl{FFP} formation since at least 10\% of them must have formed by ejection from a disc. 

The discovery of a large population of \glspl{FFP} in \gls{USC} and \gls{oph} also bears important implications on the formation and early evolution of planetary systems and, specifically, on the timescale of the processes involved. N-body simulations indeed predict that most dynamical instabilities happen within 0.1--1 Myr of the planets' formation, although there do exist configurations that produce later instabilities\cite{Chatterjee+2008,Juric+2008,Raymond+2010,vanElteren+2019}.  Our results suggest that giant planet systems must form and become dynamically unstable within the observed lifetime of the region of $3-10$~Myr to contribute to the population of \glspl{FFP}.  While the instability among the Solar System's giant planets\cite{Nesvorny2018} was much less violent than those for the mass range of \glspl{FFP} in our sample\cite{Raymond+2018}, current studies suggest that it may have also happened early.\cite{Clement+2018,Morbidelli+2018}

Instabilities can also be the result of close stellar encounters: numerical simulations have shown that dynamical interactions with other stars in clustered environments may induce instability in planetary systems or even liberate planets, enriching the \gls{FFP} population\cite{Parker+2012,vanElteren+2019}. Recent studies showing that the demographics of exoplanets depends on the stellar environment\cite{Winter+2020} confirm that such interactions must indeed play a role and contribute to the observed population of \glspl{FFP}. Our observations suggest that these encounters might take place within the first 10~Myr of a system's life.

The combined contributions of \glspl{FFP} from core-collapse (13--118\%) and ejection from planetary systems (10--130\%) derived from our analysis can explain the formation of the majority of \glspl{FFP}. But other mechanisms are known to be at work: photo-erosion of prestellar cores\cite{Hester+1996, Whitworth+2004} has been observed around massive B stars\cite{2009A&A...493..931B,2009ApJ...701L.100H}. Since \gls{USC} is an OB association, some of the \glspl{FFP} might have formed by photo-erosion. At the same time, given that photo-erosion can only occur in the direct vicinity of relatively scarce OB stars, we can reasonably assume that the contribution of this mechanism to the overall \gls{FFP} population must be fairly small.
The contribution of dynamical ejections of sub-stellar embryos in the parent prestellar cluster\cite{Bate+2002, Reipurth+2001} could in principle be significant source of \glspl{FFP}.  However, recent hydrodynamical simulations show that while such dynamical ejections can produce a realistic population brown dwarfs, they under-produce \glspl{FFP}\cite{Bate+2019} (see Figure~\ref{fig:mag-mass-func}).
This new sample of \glspl{FFP} is by far the largest and most comprehensive known to date and brings exciting opportunities to better understand their origin by performing statistically robust studies of their properties. The multiplicity, kinematics and properties of discs among \glspl{FFP} are expected to depend on the processes at work and should hold important clues on their formation and early evolution. Finally, this new sample also provides an opportunity to study the atmospheres of planetary-mass objects in the absence of a blinding host star.

%\newpage
\begin{methods}

\noindent{\bf \large DANCe catalogue}

The COSMIC-DANCe project\cite{Bouy+13} aims at performing deep and complete censuses of young nearby associations down to the planetary mass regime. While the identification of young stars and even brown dwarfs has become trivial with the advent of \textit{Gaia}, finding the extremely faint  free-floating planets well beyond Gaia's sensitivity limit remains a difficult challenge that requires deep ground-based observations. The strategy chosen for COSMIC-DANCe relies on the measurement of proper motions and multi-wavelength photometry and the use of modern data mining techniques to identify the faintest members. We therefore combine optical and infrared wide-field images found in public archives with our own observations to obtain multi-epoch and multi-wavelength catalogues of young clusters and star forming regions in the solar neighbourhood. The relatively long time base-line encompassed by this data set ($\sim20$ years) allows us to derive proper motions with a precision of the order of 1~mas~yr$^{-1}$.

We searched in the \gls{ESO}, \gls{NOIRLab}, \gls{PTF}, \gls{CADC}, \gls{ING}, \gls{WSA}, and \gls{SMOKA} public archives for wide-field images inside the area
\begin{equation}
\begin{split}
    235\degr<  & ~\textup{RA}   <252\degr, \\
   -29.5\degr< & ~\textup{Dec}  <-16.7\degr.
\end{split}
\label{eq:usc-area}
\end{equation}
We complemented the data found in these public archives with our observations with the \gls{DECam} mounted on the Blanco telescope at the \gls{CTIO}, the VISTA and VST telescopes at \gls{ESO}, the MegaCam camera at \gls{CFHT},  the NEWFIRM camera mounted on the $4$~m telescope at the \gls{KPNO} and \gls{CTIO}, the \gls{HSC} mounted on the Subaru telescope and the \gls{WFC} mounted on the \gls{INT}. Supplementary Table~1 %\ref{tab:instruments} 
gives an overview of the various instruments used for this study. We used the \textsc{MaxiMask} and \textsc{MaxiTrack} softwares\cite{maximask} to detect problematic pixels (cosmic rays, dead/hot pixels, satellite trails, saturated stars and associated blooming artefacts, diffraction spikes) and problematic astronomical images (e.g. tracking lost). Additionally, we discarded a few images after a visual inspection because of their poor quality, limited sensitivity, or acquisition problems.
Finally, we collected $80\,818$~individual images of $18$~different instruments, obtained over the past $20$~years. The raw and processed data added up to almost $120$~TB and were processed and analysed on a dedicated HPC server. We extracted over $1.3$~billion individual detections from these images. The image reduction and the photometric and astrometric analysis are described in a previous article\cite{Bouy+13}. Briefly, all images were either processed using their official pipeline (e.g. \gls{DECam}, \gls{HSC}) or using \emph{Alambic}\cite{Vandame2002} following standard procedures. Sources were detected, extracted and their astrometry and photometry measured using {\sc SExtractor}\cite{Bertin+96} and PSFEx\cite{Bertin+13}. The astrometric and photometric calibration was then obtained using SCAMP\cite{Bertin+2006}, and nightly sets of individual images were average-combined (weighted by their exposure times) using SWarp\cite{Bertin2010}. Sources were then extracted in these deeper nightly stacks, and proper motions and photometry measured using SCAMP again.  Deep stacks were also produced combining all available images obtained in a given camera+filter and used for the photometry only. The details of the whole  procedure can be found in the original COSMIC-DANCE article\cite{Bouy+13}. In Supplementary Figure~1 %\ref{fig:pm_error} 
we show the precision of proper motions as a function of magnitude. Because of \textit{Gaia} superiority compared to our ground based observations we always use the astrometry from \gls{GDR2} when available and the astrometry from DANCe elsewhere. This explains the shift in precision observed at $i\sim21$~mag.

Our catalogue contains proper motions and photometry ($grizyJHK$) for $40\,882\,164$ unique sources. To optimise the number of sources with complete photometry (essential for the membership analysis, see description of the membership analysis) and accelerate the computational time of the membership algorithm, we selected the area where the coverage of most instruments was best. We defined an elliptical area centred in (RA $=243.5$\degr, Dec $=-23.1$\degr) with a semi-major axis of $8.5\degr$ in RA and a semi-minor axis of $6.4\degr$ in Dec. This selection roughly follows the coverage of the UKIDSS near-infrared survey\cite{Lawrence+07} which we also used to define our \gls{DECam} and \gls{HSC} surveys. 

The globular cluster NGC 6121 ($\mu_\alpha^*=-12.48$~mas~yr$^{-1}$, $\mu_\delta=-18.9$~mas~yr$^{-1}$ \cite{Baumgardt+2019}) is inside the area covered by our survey. To avoid contamination from its members in our sample, we discarded the sources encompassed in a circular region of $12\arcmin$ around the globular cluster centre (RA$=245.896$\degr, Dec$=-26.527$\degr). The final catalogue contains $28\,062\,542$ sources and has a median precision of $<1$~mas~yr$^{-1}$ in proper motions for sources brighter than $i<20$~mag. Supplementary Figure~2 %\ref{fig:completeness-DANCe-USC} 
shows the density distribution of sources as a function of magnitude for different filters. We used the maximum of this distribution as the approximate completeness limit in each band. However, it depends on dust extinction and varies with position in the DANCe catalogue.

\noindent{\bf \large Membership analysis}

We used a maximum likelihood approach to infer the parameters of the models describing both the cluster and field populations\cite{Sarro+14, Olivares+19}. The algorithm models first the distribution in the space of observables (parallaxes, proper motions and photometry) of the sources that belong to the field and then iteratively searches for a maximum likelihood solution for the parameters of the model that describes the distribution of observables for the cluster sources. In each iteration, the algorithm calculates the membership probabilities using Bayes' theorem and the fractions of cluster and field members as priors. The sources with missing data cannot be used to infer the cluster and field models and, for that, it is of uttermost importance to define an adequate representation space, i.e. a set of parameters which is the largest but at the same time contains a large fraction of sources with complete observations. We note however that our algorithm uses the final model (computed with complete sources) to obtain a membership probability for the sources with partial information by marginalising over the missing information. We searched for members in three different catalogues: the DANCe catalogue (produced in this work), the \textit{Gaia} DR2 catalogue and the \textit{Hipparcos} catalogue. These catalogues include very different photometric bands, and we decided to run an independent analysis for each catalogue. 
The parameters (photometry and astrometry) used in each case are described in the following paragraphs.

\textbf{Initial members}

We compiled a list of $2\,865$~published candidate members in the literature\cite{Damiani+2019, Luhman+2018, Muzic+2012, Ducourant+2017, Lodieu+2018} in the area covered by this study. We cross-matched this list with each of our three catalogues (\textit{Hipparcos}, \textit{Gaia} and DANCe) to obtain the initial list to start each analysis. In the case of \textit{Hipparcos}, we excluded Antares ($\alpha$~Sco) since it is a giant star, and therefore, it falls out of the empirical pre-main-sequence isochrone. For the analysis with \textit{Gaia} and DANCe, we excluded the most extinguished members since they confuse our empiric isochrone. 

\textbf{Representation space}

For the analysis with \textit{Hipparcos}, we searched for members in the space of \texttt{pmra}, \texttt{pmdec}, \texttt{parallax}, $V$, $B-V$, where all the sources in the catalogue have complete observations.
For the analysis with \textit{Gaia}, we used the representation space \texttt{pmra}, \texttt{pmdec}, \texttt{parallax}, $G_{\rm RP}$, $G-G_{\rm RP}$, excluding the $G_{BP}$ band which is less accurate for cool dwarfs\cite{Maiz-Apellaniz+2018}. In this space, $7\,768\,856$~sources ($97\%$) have complete observations.
For the analysis with the DANCe catalogue, we used the representation space \texttt{pmra}, \texttt{pmdec}, $i$, $J$, $H$, $i-Ks$. We combined the $i$ band in the optical, which has the largest coverage, with the infrared bands $J, H, Ks$ where the ultracool dwarfs are best detected. With this representation space, $10\,483\,667$~sources have observations in all the photometric bands, which represents $37$\% of the catalogue. We decided not to include the \textit{g}, \textit{r}, \textit{z}, and \textit{$Y$} bands in the representation space because they reduced the number of sources with complete photometry, specially for the coolest objects. 

\textbf{Field model}

The model of the field population is a \gls{GMM} in the whole representation space. We explored models with different number of Gaussians and used the \gls{BIC} criterion to chose the final model. Since the \textit{Hipparcos} catalogue has a reduced number of sources, we explored models with several components between $1$ and $20$ and selected six as the optimum choice according to the \gls{BIC}. For the \textit{Gaia} catalogue, we explored models between $20$ and $180$~Gaussians and chose $60$ as the optimum choice according to the \gls{BIC}. Finally, for the DANCe catalogue, we explored models between $60$ and $300$ and chose $100$~components.

\textbf{Cluster model}

The proper motion distribution of the region of \gls{USC} and \gls{oph} is much more complex than that of open clusters previously analysed with this methodology\cite{Sarro+14, Olivares+19, Miret-Roig+19}. While open clusters have symmetric nearly Gaussian distributions in astrometry, this young region shows a rich substructure far from Gaussian (see Supplementary Figure~3%\ref{fig:VPD-USC}
) and indicative of multiple kinematic populations. To model this complex distribution, we used a \gls{GMM} where the Gaussians are not necessarily concentric and explored models with between $1$ and $10$ Gaussians. Since the \textit{Hipparcos} catalogue contains a very reduced number of sources, we found that a single multivariate Gaussian function suffices to model the cluster proper motions and parallaxes. The \textit{Gaia} and DANCe catalogues are much larger and the number of Gaussians selected according to the \gls{BIC} criterion is $5-7$ for \textit{Gaia} and $5-6$ for DANCe (depending on the parameter $p_{in}$, see below).

We ran the model with different internal probability thresholds\cite{Sarro+14}, i.e. different degrees of completeness and contamination ($p_{in} = 0.5, 0.6, 0.7, 0.8$, and $0.9$), and for each we computed the optimum threshold, $p_{opt}$, using synthetic data\cite{Sarro+14, Olivares+19}. In Supplementary Table~2%\ref{tab:pin-usc}
, we show  $p_{in}$, $p_{opt}$, and the number of members for each independent analysis (\textit{Hipparcos}, \gls{GDR2}, and DANCe). 

\textbf{Final list of members}

Membership probabilities obtained from the analysis with different $p_{in}$ values, as well as the astrometry and photometry used in the \textit{Hipparcos}, \textit{Gaia}, and DANCe catalogues, are available at \gls{CDS}. Choosing the best solution (the best $p_{in}$) is a non-trivial decision, which depends on the aim of the study. To study the magnitude distribution and mass function, we need a list of members as complete as possible. For this reason, we prefer solutions with low $p_{in}$ values which have greater completeness, although they can also be slightly more contaminated.

First, we compared the \textit{Gaia} solutions obtained with different $p_{in}$ and found that $2\,603$ sources ($94\%$) are the same in all the lists. Additionally, the contamination rate and true positive rate computed with synthetic data are very similar in the five studies (see Supplementary Table~2%\ref{tab:pin-usc}
), so we had no prior reason to prefer one list to another. Therefore, we chose the list of $p_{in}=0.5$ as the final list of \textit{Gaia} since it was the one with the largest number of members. Following an analogous procedure with the DANCe solution, we also chose the list of $p_{in}=0.5$. With the \textit{Hipparcos} study, we selected the solution of $p_{in}=0.7$, which represents a good compromise between low contamination and high completeness. To this final list, we added the giant star Antares manually. Our final list of members contains 3\,455 sources from the \textit{Hipparcos}, \gls{GDR2}, and DANCe catalogues. 

\textbf{Membership completeness}

The completeness of our membership analysis depends on the completeness of the astro-photometric catalogue and the membership algorithm. Our optical \gls{DECam} and near-infrared VISTA images and archival UKIRT images cover the entire area and ensure that the instrumental $i,Y,J,H,Ks$ sensitivities are fairly homogeneous spatially. To get an estimate of the completeness in the substellar mass regime, we propagated the apparent magnitude completeness of the DANCe catalogue to masses. The limiting magnitude to search for ultra-cool dwarfs is set by the $i$ band, which we estimated to be sensitive up to $i\sim26$~mag and complete up to $i\sim23$~mag (see Supplementary Figure~2%\ref{fig:completeness-DANCe-USC}
). This approximate magnitude limit completeness corresponds to masses between 7~M$_{Jup}$ (assuming an age of 3~Myr) and 13~M$_{Jup}$ (assuming an age of 10~Myr). However, our membership algorithm is expected to miss highly extincted objects ($A_{\rm V}\gtrsim3$~mag) and sources with near-infrared excess related to the presence of circumstellar material.

\textbf{Membership validation}

The membership classification mostly coincides in the \textit{Hipparcos}--\textit{Gaia} and \textit{Gaia}--DANCe studies, in the magnitude range where both catalogues are complete. The small differences between catalogues can be attributed to the different information provided by each of them. We used the \textit{Gaia} membership analysis, with the additional information on the parallax, to estimate a contamination rate of $8\%$ on the DANCe membership in the magnitude range $9<J<14$~mag assuming that the \textit{Gaia} selection is perfectly clean. Similarly, the comparison between the \textit{Gaia} and DANCe samples over the common luminosity domain shows that  one third of the objects identified with \textit{Gaia} are not recovered with DANCe because of either missing photometry, high extinction or near-infrared excess likely related to the presence of a circumstellar disc. Therefore, the completeness of our census is expected to be better in \gls{USC} than in \gls{oph} since extinction is much lower and near-infrared excesses related to discs should be less frequent given the more advanced age and timescale for disc decay. 

We note that at $J\sim10$~mag (see Fig.~\ref{fig:CMD}) there are some contaminants. These are sources identified with the DANCe membership analysis, therefore, using only the proper motions and photometry (no parallaxes). The proper motions are compatible with the proper motion distribution of the association and that is why they are classified as members. Additionally, since they are bright stars, their photometry saturates in many bands so we have little photometric information. In any case, these represent a 1\% of the members which is within the contamination rate that we estimated. 

We recovered the majority of members previously reported in the literature\cite{Damiani+2019, Lodieu+2018, Muzic+2012, Ducourant+2017, Esplin+2018}. Supplementary Figure~4 %\ref{fig:lit-members} 
shows a comparison between the members reported by previous studies and the members found in this study, using the photometry and proper motions we measured in this work. In this Figure, we only considered studies sensitive to \glspl{FFP}. We missed around 80 substellar members previously reported in the literature, most of which are in \gls{oph}, are highly extincted or host circumstellar discs (as can be seen in the colour-magnitude diagrams). Besides, some of the members reported in the literature are discarded by our membership analysis because of their inconsistent proper motions. This is especially significant in a recent study\cite{Lodieu+2021} where only half of their members are identified in our astro-photometric analysis. The other half are either classified as non-members (having photometric and/or proper motion measurements inconsistent with the association) or undetected.
In this study, we add $\sim$800 of new members, 70--170 of which are \glspl{FFP}, depending on the age assumed.

\noindent{\bf \large The mass function}

\textbf{Distances}

We used \textit{Kalkayotl}\footnote{\url{https://github.com/olivares-j/kalkayotl}}~\cite{Olivares+2020} to infer Bayesian distances for all the members with a parallax measurement in the \gls{GDR2} catalogue. We used a Gaussian prior with a locus and scale of 145~pc and 45~pc, corresponding to the median and three times the standard deviation of the distribution of distances obtained inverting the parallax. For the sources in the DANCe catalogue, without a parallax measurement, we sampled the distance from the cluster distance distribution obtained with all the \textit{Gaia} members. 

\textbf{Masses}

We combined the apparent photometry ($grizyJHKs$) and the distance estimate of each star to obtain absolute magnitudes. These were compared, in a Bayesian framework, to theoretical evolutionary models to infer the posterior distribution of the mass and extinction of each source with \textit{Sakam}\footnote{\url{https://github.com/olivares-j/Sakam}}~\cite{Olivares+19}. This algorithm ignores any possible source of error related to the theoretical evolutionary model chosen by the user. The model does not include effects on the variability of the source due to binarity, activity, or other factors. These effects eventually end up included in the extinction estimate, enlarging its uncertainties. Finally, the mass and the extinction are degenerated but at least half of the planetary-mass objects we found are in regions of low extinction, favouring their planetary nature.

We had to use different models for the high and low-mass regimes as there is no single set of models covering the entire mass range of our members. We combined the PARSEC-COLIBRI\footnote{\url{http://stev.oapd.inaf.it/cgi-bin/cmd}}~ models\cite{Marigo+17} and BHAC$15$\footnote{\url{http://perso.ens-lyon.fr/isabelle.baraffe/}}~ models\cite{Baraffe+15} which cover the high and low-mass range of our members, respectively. We find that both grids of models agree fairly well around 0.5~M$_\sun$ and decided to use the masses inferred from the BHAC$15$ models  below 0.5~M$_\sun$ and the masses inferred with the PARSEC-COLIBRI models above 0.5~M$_\sun$. In Supplementary Figure~5 %\ref{fig:SED} 
we show two examples of the best-fit spectral energy distribution models obtained with \textit{Sakam} for a brown dwarf and a planetary-mass object. The complete final list of members with the masses and extinctions inferred with \textit{Sakam} is available at the \gls{CDS}.

\textbf{Magnitude and mass distributions}

To obtain the magnitude (mass) distribution, we sampled the individual magnitude (mass) of each source with a Gaussian centred at the measured magnitude (mass) and a standard deviation equal to the uncertainty. Then, we defined a grid between the least and most bright (massive) object in our sample and added the contribution of all the sources to each magnitude (mass) bin. We convoluted this distribution with a Gaussian \gls{KDE} with a bandwidth chosen according to Scott's\cite{Scott+92} and  Silverman's\cite{Silverman+86} rules. We estimated the uncertainties in the magnitude (mass) function with a bootstrap of $100$~repetitions and reported the $1\sigma$ and $3\sigma$ confidence levels.

 \textbf{Data availability} The data that support the findings of this study will be available at the CDS after the reviewing process.

\end{methods}

\pagebreak
%\newpage

\newpage
\begin{addendum}

 \item We are grateful to Paolo Padoan, Matthew Bate and Veli-Matti Pelkonen for insightful comments on the comparison of our observational mass function to simulations, to Andrew Howard, Gijs Mulders, Christophe Lovis for input on the occurrence rates and to Karla Peña Ramírez, Alexander Scholz, Nicolas Lodieu for input on FFPs in star forming regions. We thank two anonymous reviewers for helpful comments.

 This research has received funding from the European Research Council (ERC) under the European Union’s Horizon 2020 research and innovation programme (grant agreement No 682903, P.I. H. Bouy), and from the French State in the framework of the ”Investments for the future” Program, IdEx Bordeaux, reference ANR-10-IDEX-03-02. H. Bouy acknowledges financial support from the Canon Foundation in Europe. S.N.R acknowledges support from the CNRS's PNP program. 
 This research has been funded by the Spanish State Research Agency (AEI) Projects  PID2019-107061GB-C61 and No. MDM-2017-0737 Unidad de Excelencia “María de Maeztu”- Centro de Astrobiología (CSIC/INTA).

  We gratefully acknowledge the support of NVIDIA Corporation with the donation of one of the Titan Xp GPUs used for this research.
  Based on observations made with the INT operated on the island of La Palma by the Isaac Newton Group in the Spanish Observatorio del Roque de los Muchachos of the Instituto de Astrofísica de Canarias. 
  This paper makes use of data obtained from the Isaac Newton Group Archive which is maintained as part of the CASU Astronomical Data Centre at the Institute of Astronomy, Cambridge.
  Based on data obtained from the ESO Science Archive Facility and with ESO Telescopes at the La Silla Paranal Observatory under programme ID 065.I-0003,065.I-0004,065.L-0463,065.O-0298,071.A-9007(A),071.A-9011(A),075.C-0419(A),075.D-0111(A),075.D-0662(C),079.A-9202(A),079.A-9203(A),079.A-9208(A),079.D-0782(A),079.D-0918(A),080.A-9210(A),081.A-9200(A),081.A-9211(A),081.A-9212(A),082.A-9212(A),082.C-0946(B),083.A-9021(A),083.A-9202(A),083.A-9204(A),083.C-0446(A),085.A-9008(A),085.A-9011(A),085.C-0690(B),085.D-0143(A),086.C-0168(D),088.C-0434(A),091.A-0507(A),091.C-0454(A),093.A-9028(B),094.A-9028(C),096.A-9021(A),097.A-9020(A),097.A-9025(C),164.O-0561(F),60.A-9120(A),67.A-0403(A),68.D-0002(B),68.D-0265(A),69.A-0615(B),69.C-0182(A),69.C-0260(A),69.C-0426(C),69.D-0582(A),71.C-0580(A),71.C-0580(B),71.D-0014(A),081.A-0673(A),083.A-0321(A),085.C-0841(E),085.C-1009(A),089.C-0952(B),089.C-0952(C),089.C-0952(E),089.D-0291(A),091.A-0703(B),091.C-0543(B),091.C-0543(C),091.C-0543(D),091.C-0543(E),092.C-0548(F),195.B-0283(A),60.A-9283(A),60.A-9800(L),60.A-9800(H),083.C-0556(A),279.C-5062(C),093.B-0280(B),095.D-0494(A),096.C-0730(A),097.C-0749(A),098.C-0850(A),099.C-0474(A),177.D-3023(G),60.A-9038(A),088.D-0675(A),089.C-0102(A),089.C-0102(B),089.C-0102(C),095.D-0038(A),097.C-0781(A),179.A-2010(H),179.A-2010(J),179.A-2010(K),179.A-2010(L),179.A-2010(N),198.C-2009(A),198.C-2009(B),198.C-2009(F),198.C-2009(H),198.C-2009(I),60.A-9292(A). 
  This work is based on data obtained from the VISIONS ESO Public survey (\url{https://visions.univie.ac.at/}).
  This project used data obtained with the Dark Energy Camera (DECam), which was constructed by the Dark Energy Survey (DES) collaboration. Funding for the DES Projects has been provided by the DOE and NSF (USA), MISE (Spain), STFC (UK), HEFCE (UK), NCSA (UIUC), KICP (U. Chicago), CCAPP (Ohio State), MIFPA (Texas A\&M), CNPQ, FAPERJ, FINEP (Brazil), MINECO (Spain), DFG (Germany) and the Collaborating Institutions in the Dark Energy Survey, which are Argonne Lab, UC Santa Cruz, University of Cambridge, CIEMAT-Madrid, University of Chicago, University College London, DES-Brazil Consortium, University of Edinburgh, ETH Zürich, Fermilab, University of Illinois, ICE (IEEC-CSIC), IFAE Barcelona, Lawrence Berkeley Lab, LMU München and the associated Excellence Cluster Universe, University of Michigan, NOIRLab, University of Nottingham, Ohio State University, OzDES Membership Consortium, University of Pennsylvania, University of Portsmouth, SLAC National Lab, Stanford University, University of Sussex, and Texas A\&M University.
Based on observations at Cerro Tololo Inter-American Observatory atNSF’s NOIRLab (NOIRLab Prop. 2012B-0569 (PI: Allen); 2013A-0214 (PI: Berger); 2013A-0327 (PI: Rest); 2013A-0351 (PI: Dey); 2013A-0723 (PI: Mamajek); 2013A-0737 (PI: Sheppard); 2013A-9999 (PI: Walker); 2013B-0325 (PI: Vivas); 2013B-0531 (PI: Mamajek); 2013B-0536 (PI: Allen); 2014A-0035 (PI: Bouy); 2014A-0239 (PI: Sullivan); 2014A-0306 (PI: Dai); 2014A-0327 (PI: Rest); 2014A-0412 (PI: Rest); 2014A-0479 (PI: Sheppard); 2014A-0480 (PI: Rich); 2014A-0634 (PI: James); 2015A-0151 (PI: Calamida); 2015A-0205 (PI: Mamajek); 2015A-0371 (PI: Rest); 2015A-0397 (PI: Rest); 2015A-0610 (PI: Fuentes); 2015B-0307 (PI: Rest); 2016A-0189 (PI: Rest); 2016A-0327 (PI: Finkbeiner); 2016B-0279 (PI: Finkbeiner); 2016B-0301 (PI: Rest); 2016B-0301 (PI: Zenteno); 2017A-0002 (PI: Bouy); 2017A-0388 (PI: Zenteno); 2017A-0389 (PI: Rest); 2017A-0389 (PI: Tucker); 2017A-0918 (PI: Yip); 2017B-0285 (PI: Rest); 2018A-0059 (PI: Bouy); 2018A-0251 (PI: Finkbeiner); 2019A-0060 (PI: Bouy); 2019A-0101 (PI: Hartigan); 2019A-0305 (PI: Drlica-Wagner); 2019A-0337 (PI: Trilling); 2019B-0323 (PI: Zenteno)), which is managed by the Association of Universities for Research in Astronomy (AURA) under a cooperative agreement with the National Science Foundation.
Based in part on observations at Kitt Peak National Observatory at NSF’s NOIRLab (NOIRLab Prop. ID 2013A-0399 (PI: Adam); 2014A-0642 (PI: Ronald)), which is managed by the Association of Universities for Research in Astronomy (AURA) under a cooperative agreement with the National Science Foundation. The authors are honored to be permitted to conduct astronomical research on Iolkam Du’ag (Kitt Peak), a mountain with particular significance to the Tohono O’odham.
Based in part on observations made at Cerro Tololo Inter-American Observatory at NSF’s NOIRLab (NOIRLab Prop. ID 2011A-0368 (PI: Stringfellow); 2010A-0475 (PI: Stringfellow); 2011A-0603 (PI: Tilvi); 2011A-0644 (PI: Catelan); 2010A-0036 (PI: Probst); 2005A-0183 (PI: Ridge); 2006A-0139 (PI: Ridge); 2006B-0021 (PI: Grindlay); 2007A-0180 (PI: Martin); 2007A-0514 (PI: Mardones); 2007A-0599 (PI: Huard); 2008B-0368 (PI: Zuckerman); 2008B-0909 (PI: Vaduvescu); 2010A-0260 (PI: Faherty); 2010A-0326 (PI: Allers); 2010A-0482 (PI: Miller)r), which is managed by the Association of Universities for Research in Astronomy (AURA) under a cooperative agreement with the National Science Foundation.
  This research used the facilities of the Canadian Astronomy Data Centre operated by the National Research Council of Canada with the support of the Canadian Space Agency.
  Based in part on data collected at Subaru Telescope which is operated by the National Astronomical Observatory of Japan and obtained from the SMOKA, which is operated by the Astronomy Data Center, National Astronomical Observatory of Japan.
  The Hyper Suprime-Cam (HSC) collaboration includes the astronomical communities of Japan and Taiwan, and Princeton University. The HSC instrumentation and software were developed by the National Astronomical Observatory of Japan (NAOJ), the Kavli Institute for the Physics and Mathematics of the Universe (Kavli IPMU), the University of Tokyo, the High Energy Accelerator Research Organization (KEK), the Academia Sinica Institute for Astronomy and Astrophysics in Taiwan (ASIAA), and Princeton University. Funding was contributed by the FIRST program from Japanese Cabinet Office, the Ministry of Education, Culture, Sports, Science and Technology (MEXT), the Japan Society for the Promotion of Science (JSPS), Japan Science and Technology Agency (JST), the Toray Science Foundation, NAOJ, Kavli IPMU, KEK, ASIAA, and Princeton University. 
  Based on observations obtained with MegaPrime/MegaCam, a joint project of CFHT and CEA/DAPNIA, at the Canada-France-Hawaii Tele
scope (CFHT) which is operated by the National Research Council (NRC) of Canada, the Institut National des Science de l'Univers of the
Centre National de la Recherche Scientifique (CNRS) of France, and the University of Hawaii.
Based on CFH12K observations obtained at the Canada-France-Hawaii Telescope (CFHT).
Based on observations obtained with WIRCam, a joint project of CFHT, Taiwan, Korea, Canada, France, and the Canada-France-Hawaii T
elescope (CFHT).
  This work has made use of data from the European Space Agency (ESA) mission {\it Gaia} (\url{https://www.cosmos.esa.int/gaia}), processed by the {\it Gaia} Data Processing and Analysis Consortium (DPAC, \url{https://www.cosmos.esa.int/web/gaia/dpac/consortium}). Funding for the DPAC has been provided by national institutions, in particular the institutions participating in the {\it Gaia} Multilateral Agreement.
  This publication makes use of data products from the Two Micron All Sky Survey, which is a joint project of the University of Massachusetts and the Infrared Processing and Analysis Center/California Institute of Technology, funded by the National Aeronautics and Space Administration and the National Science Foundation.
This paper makes use of software developed for the Large Synoptic Survey Telescope. We thank the LSST Project for making their code available as free software at  http://dm.lsst.org
The Pan-STARRS1 Surveys (PS1) have been made possible through contributions of the Institute for Astronomy, the University of Hawaii, the Pan-STARRS Project Office, the Max-Planck Society and its participating institutes, the Max Planck Institute for Astronomy, Heidelberg and the Max Planck Institute for Extraterrestrial Physics, Garching, The Johns Hopkins University, Durham University, the University of Edinburgh, Queen’s University Belfast, the Harvard-Smithsonian Center for Astrophysics, the Las Cumbres Observatory Global Telescope Network Incorporated, the National Central University of Taiwan, the Space Telescope Science Institute, the National Aeronautics and Space Administration under Grant No. NNX08AR22G issued through the Planetary Science Division of the NASA Science Mission Directorate, the National Science Foundation under Grant No. AST-1238877, the University of Maryland, and Eotvos Lorand University (ELTE) and the Los Alamos National Laboratory.
Based on data collected at the Subaru Telescope and retrieved from the HSC data archive system, which is operated by Subaru Telescope and Astronomy Data Center at National Astronomical Observatory of Japan.
Based on observations obtained with Planck (\url{http://www.esa.int/Planck}), an ESA science mission with instruments and contributions directly funded by ESA Member States, NASA, and Canada.

 \item [Author Contributions] 
 N. Miret-Roig and H. Bouy led the observations, data analysis and scientific analysis. H. Bouy is the P.I. of the COSMIC-DANCE project. S. N. Raymond contributed to the scientific analysis and the discussion about planet ejection and formation. M. Tamura led the Subaru observations used in this study. E. Bertin wrote the software packages used to process and analyse the images. J.-C. Cuillandre, P. Galli, D. Barrado and N. Huélamo led the observations at various observatories. J. Olivares, L. Sarro, A. Berihuete led the development of the probabilistic method and software used to identify members. 
 
 \item[Competing Interests] The authors declare that they have no
competing financial interests.

 \item[Correspondence] Correspondence and requests for materials should be addressed to N.M.R (\url{nuria.miret.roig@univie.ac.at}).

\end{addendum}

\newpage

%%
%% TABLES
%%
\begin{table}
    \begin{center}
    \renewcommand{\arraystretch}{1.5}
    \caption{Planetary occurrence rate for different studies.}
    \begin{tabular}{|c||c|c|c|c||c|c|}
    \hline
    \hline%\\[0.2cm]
    Method & POR (\%) & Mass   & Separation  & SpT star & $f_\textup{FFP ejected}$ & $\frac{f_\textup{FFP ejected}}{f_\textup{FFP observed}}$\\[0.2cm]
    \hline
    \hline

    RV\cite{Fernandes+2019, Mayor+2011} & $2.1\pm0.5$ & 4-13~M$_J$ & 0.1-100~AU & solar type & 0.016-0.021 & 23--130\% \\
    \hline 
    DI\cite{Bowler2016}     & $0.6^{+0.7}_{-0.5}$    & 5-13~M$_J$  & 30--300~AU & BAFGKM   & 0.005-0.006 & 7--38\% \\
    DI\cite{Baron+2019}     & $1.83^{+5.76}_{-0.62}$ & 5-13~M$_J$  & 30--300~AU & BAFGKM   & 0.014-0.018 & 20--115\% \\
    \hline 
    ML\cite{Suzuki+2016}   & $1-2$             & 0.007--0.02$^*$ & 0.2--5~$\Theta_E$ & all & 0.008-0.02 & 11--125\% \\   
    \hline 
    RV+DI+ML\cite{Clanton+2016}   & $\sim4$   & 1-13~M$_J~^\dag$  & 2--1000~AU & all & 0.03-0.04 & 44--250\% \\   
    \hline
    \hline

   \end{tabular}
     \label{tab:POR}
  \end{center}
   \noindent Col.~1: planet detection method: radial velocity (RV), direct imaging (DI), microlensing (ML), Col.~2: planet occurrence rate (POR), Col.~3: mass range of the planets, Col.~4: separation or semimajor axis, Col.~5: spectral types of the primary body, Col.~6: estimated fraction of ejected FFPs using Eq.~\ref{eq:FFP} and assuming $f_\textup{unstable}=0.75-1$ and $n_\textup{ejected}=1$ (see main text), Col.~7: estimated percentage of observed FFPs which were ejected from planetary systems. The conservative ranges we provide include both the uncertainties from our observations and current knowledge of planetary systems.\\
   \noindent $^*$ mass ratio between the planet and star mass ($q$).\\
   \noindent $^\dag$ our observations are not sensitive to masses between 1--4~M$_J$ and thus, the POR is overestimated with respect to our observations.
\end{table}

%% FIGURES

\section*{Supplementary material}

# Supplementary Information

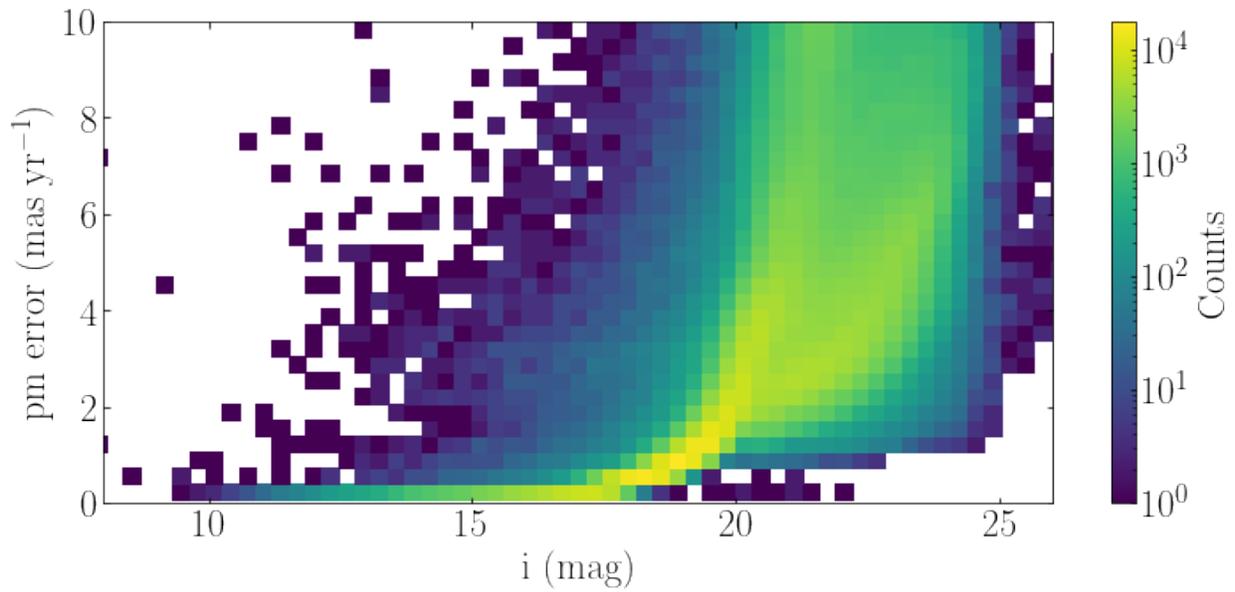

Supplementary Figure 1: Estimated proper motion error as a function of $i$ magnitude for the DANCe catalogue. We note that whenever there is a proper motion measure from *Gaia* we use it. This explains the change in precision observed at $i \sim 21$ mag.

1

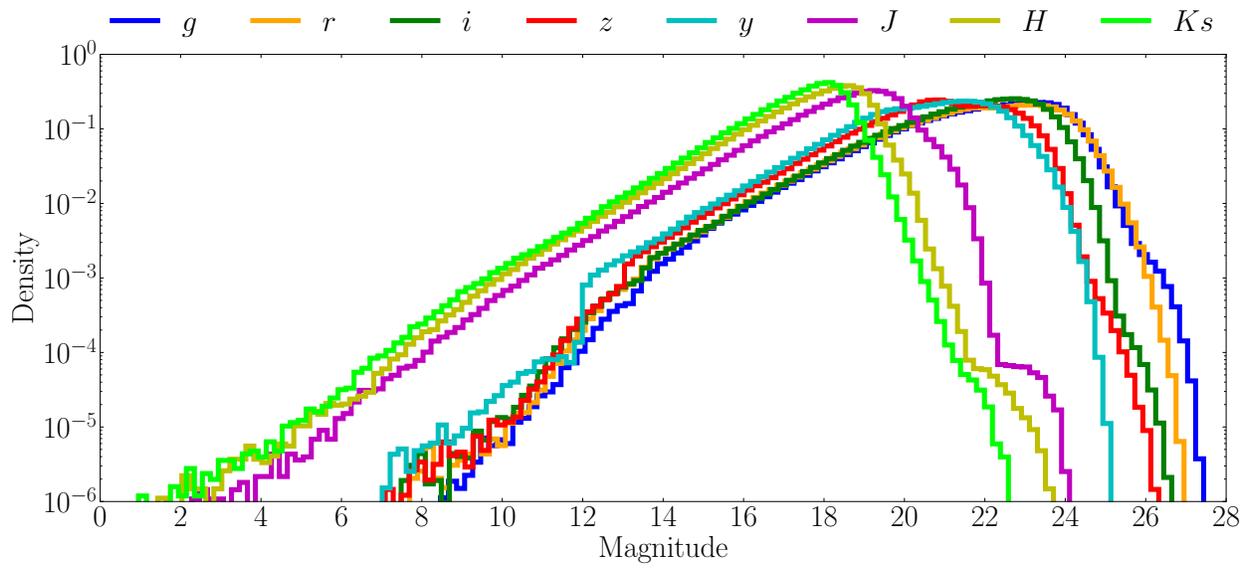

Supplementary Figure 2: Density of sources per 0.2 magnitude bin as a function of magnitude for all the sources in the DANCe catalogue.

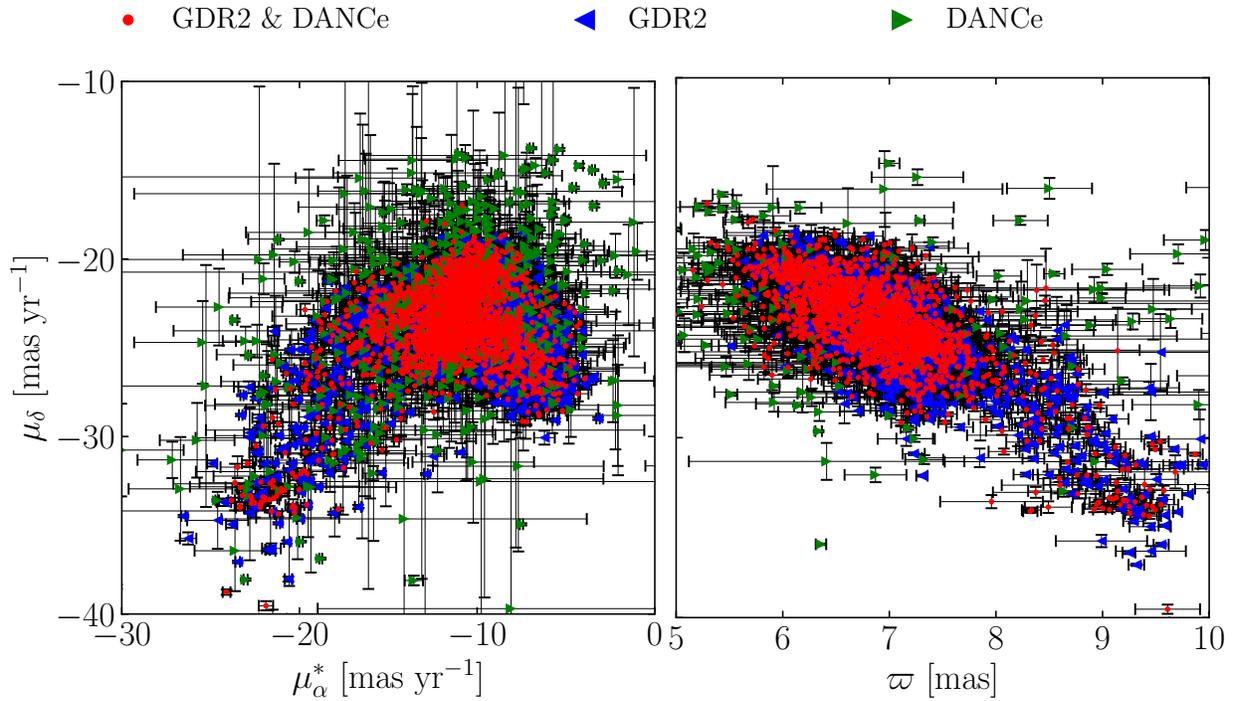

Supplementary Figure 3: Vector point diagram (left) and parallax–proper motion diagram (right) of USC and Oph. The members are shape- and colour-coded according their origin: *Gaia* and DANCe analysis (red circles), only *Gaia* analysis (blue left-pointing triangle), and only DANCe analysis (green right-pointing triangle). The error bars represent the astrometric uncertainties reported in the *Gaia* and DANCe catalogues.

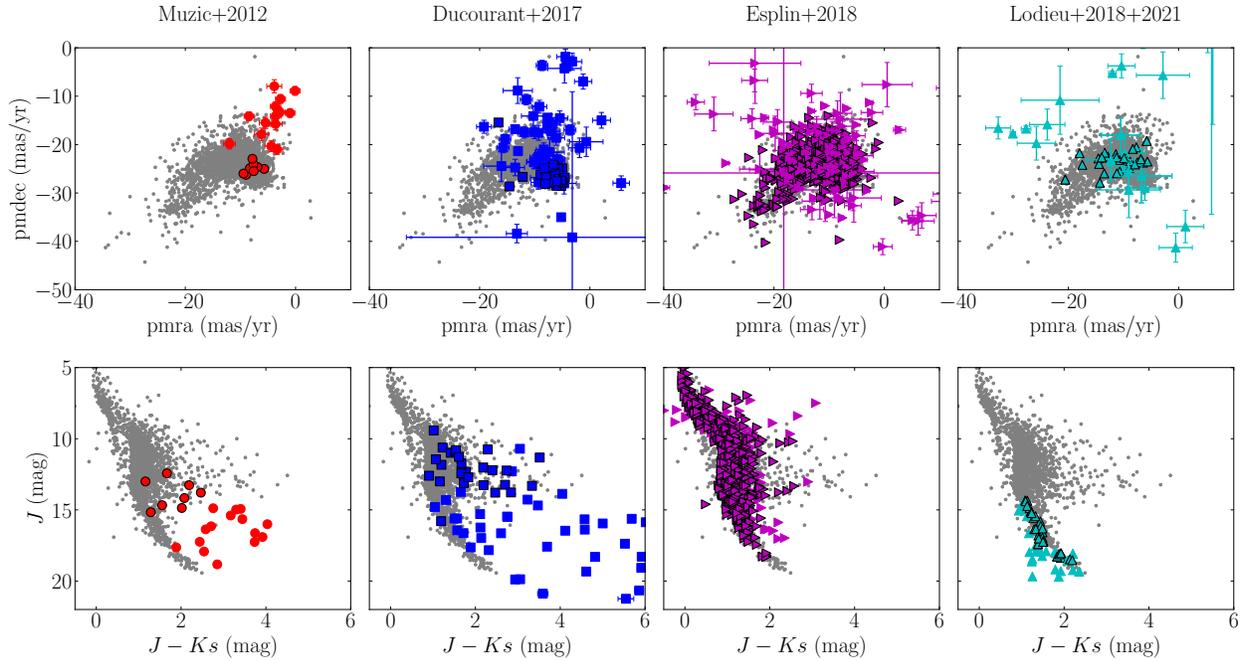

Supplementary Figure 4: Vector point diagrams (top) and colour-magnitude diagrams (bottom) of the members found in this study (grey dots) and the members reported in previous studies[1–5] (coloured markers). The members in common are indicated by a black edge line. The error bars represent the observational uncertainties in the *Gaia* and DANCe catalogues for the members previously identified in the literature and discarded by our membership analysis. We note that our study is not sensitive to highly extinct areas or the detection of sources with moderate/significant near-infrared excess.

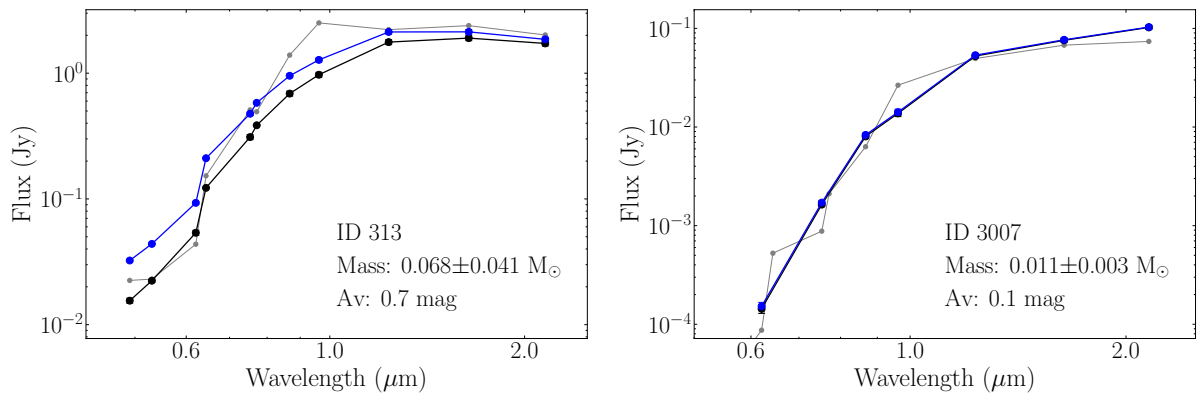

Supplementary Figure 5: Spectral energy distribution of two of our substellar objects: a brown dwarf mass object (left) and a planetary-mass object (right). The photometric observations (black) and the dereddened observations (blue) are indicated. The error bars represent the photometric uncertainties and are smaller than the markers in most cases. The BHAC15 models at 5 Myr used for the best-fit models are shown (gray).

Supplementary Table 1: Instruments used in the DANCe catalogue of USC and Oph.

| Telescope | Instrument | Filters | Platescale [pixel$^{-1}$] | Field of view | Epoch Min./Max. | Images |
|---|---|---|---|---|---|---|
| ESO VISTA[6] | VIRCAM | $z, J, H, K$ | 0″.34 | $1.2° \times 1.1°$ | 2010–2018 | 18 598 |
| ESO VST[7] | OmegaCAM | $u, g, r, i, z, H\alpha$ | 0″.21 | $1° \times 1°$ | 2014–2017 | 3302 |
| ESO VLT[8] | VIMOS | $R, I, Z$ | 0″.205 | $28' \times 32'$ | 2007–2011 | 261 |
| ESO VLT[9] | HAWK-I | $y, J, H, K$ [a] | 0″.106 | $7.5' \times 7.5'$ | 2008–2015 | 2752 |
| ESO (2.2 m)[10] | WFI | $B, V, R, I$ [a] | 0″.24 | $34' \times 33'$ | 2000–2017 | 1562 |
| CTIO (Blanco)[11] | DECam | $u, g, r, i, z, y$ [a] | 0″.27 | $1.1°$ radius | 2012–2018 | 3744 |
| CTIO (Blanco)[12] | ISPI | $J, H, K$ [a] | 0″.3 | $10.25' \times 10.25'$ | 2005–2010 | 2214 |
| CTIO (Blanco)[13] | NEWFIRM | $J, H, K$ [a] | 0″.4 | $28' \times 28'$ | 2010–2011 | 1348 |
| KPNO (Mayall)[13] | NEWFIRM | $J, H, K$ [a] | 0″.4 | $28' \times 28'$ | 2013–2014 | 247 |
| CFHT[14] | MegaCam | $u, g, r, i$ [a] | 0″.18 | $1° \times 1°$ | 2004–2016 | 1395 |
| CFHT[15] | WIRCam | $y, J, H, K$ [a] | 0″.3 | $20' \times 20'$ | 2007–2017 | 6579 |
| CFHT[16] | CFH12K | $B, V, r, i, z$ | 0″.21 | $42' \times 28'$ | 2000–2002 | 438 |
| INT[17] | WFC | $B, V, R, I, Z, g, r, i$ [a] | 0″.33 | $34' \times 34'$ | 2000–2014 | 430 |
| UKIRT[18] | WFCAM | $z, y, J, H, K$ [a] | 0″.4 | $40' \times 40'$ [b] | 2005–2012 | 34 837 |
| SDSS[19] | SDSS Imaging Camera | $u, g, r, i, z$ | 0″.396 | | 2005–2009 | 2112 |
| Subaru[20] | HSC | $r, i, Y$ | 0″.17 | $1.8°$ radius | 2015–2017 | 160 |
| Subaru[21] | Suprime-Cam | $g, r, i, z, V, R, I$ [a] | 0″.2 | $34' \times 27'$ | 2004–2011 | 804 |
| Palomar 48″[22] | PTF | $g, r$ | 1″.0 | $3°.3 \times 2°.2$ [c] | 2010–2014 | 35 |

[a] as well as narrow and medium bands
[b] the chip layout has large gaps between detectors, and the coverage of the focal plane is only partial
[c] one of the 12 detectors is dead

6

7

Supplementary Table 2: Performance of the membership analysis.

| | Hipparcos | | | | Gaia | | | | DANCe | | | |
|---|---|---|---|---|---|---|---|---|---|---|---|---|
| $p_{in}$ | $p_{opt}$ | Memb | CR (%) | TPR (%) | $p_{opt}$ | Memb | CR (%) | TPR (%) | $p_{opt}$ | Memb | CR (%) | TPR (%) |
| 0.5 | 0.97 | 116 | 17 | 92 | 0.95 | 2762 | 1.1 | 99.2 | 0.77 | 2556 | 2 | 98 |
| 0.6 | 0.96 | 108 | 7 | 95 | 0.96 | 2698 | 0.9 | 99.2 | 0.78 | 2458 | 2 | 99 |
| 0.7 | 0.94 | 112 | 7 | 96 | 0.96 | 2678 | 1.0 | 99.4 | 0.83 | 2342 | 1.5 | 99 |
| 0.8 | 0.95 | 103 | 5 | 96 | 0.96 | 2661 | 0.9 | 99.6 | 0.88 | 2185 | 0.9 | 99 |
| 0.9 | 0.96 | 78 | 3 | 98 | 0.96 | 2623 | 0.9 | 99.6 | 0.83 | 2086 | 0.9 | 99 |

Performance of the membership analysis obtained with different internal probability thresholds ($p_{in}$) for the three catalogues considered, namely Hipparcos, Gaia, and DANCe. For each internal probability threshold ($p_{in}$) we show the corresponding optimum probability threshold ($p_{opt}$), number of members (Memb), contamination rate (CR) and true positive rate (TPR). The $p_{opt}$, CR, and TPR were obtained with synthetic data (see text).



\end{document}